\documentstyle[aps,epsfig]{revtex}

\begin{document}

\title
{Spin-Boson Hamiltonian and Optical Absorption 
of Molecular Dimers} 
\author{Christiane Koch\footnote{Present address:
Fritz-Haber-Institut der Max-Planck-Gesellschaft, 
Abt. Chemische Physik, Faradayweg 4-6, 14195 Berlin, Germany} 
and Bernd Esser\footnote{Corresponding author. Fax 49 30 2093 7638;
 e-mail: bernd.esser@physik.hu-berlin.de}}

\address{Institut f\"ur Physik, Humboldt-Universit\"at zu Berlin, 
  Invalidenstr. 110,\\
  D-10115 Berlin, Germany}

\date{\today}

\maketitle

\thispagestyle{empty}

\begin{abstract}
An analysis of the eigenstates of a symmetry-broken 
spin-boson Hamiltonian is performed by computing 
Bloch and Husimi projections.
The eigenstate analysis is combined 
with the calculation of absorption bands of 
asymmetric dimer configurations constituted by monomers
with nonidentical excitation energies and optical
transition matrix elements.   
Absorption bands with regular and irregular fine
structures are obtained and related to 
the transition from the coexistence 
to a mixing of adiabatic branches in the spectrum.  
It is shown that correlations between spin
states allow for an interpolation between absorption
bands for different optical asymmetries.

\end{abstract}


\section{Introduction}

The spectral properties of the spin-boson Hamiltonian 
have frequently been adressed in the past and in the last years 
investigations on 
the spectral features of classical nonintegrability 
in the quantum case were performed (see e.g. \cite{MSLN,SE97,ST,SSKR} 
and references therein). 
The interest in the spectral and dynamical properties 
of the spin-boson Hamiltonian
is due to its numerous applications which include a large variety
of phenomena in molecular and solid state physics with one  
realization being the optical and transfer properties of 
excitonic-vibronic coupled dimers. Thus the spectral and related
optical properties of the corresponding dimer Hamiltonian have
intensively been investigated over the years, 
as examples we point to the work exposed in 
\cite{BH,SEW}. 
Most of these investigations were based on a Hamiltonian 
corresponding to a symmetric dimer configuration. 
However, from the point of view of a general asymmetric situation 
the symmetric case is singular, e.g. the quantum states possess 
parity properties in the symmetric case, 
which are not present in general.
It should be noted that dimers
often constitute subunits of more complicated
molecular aggregates with an asymmetric structure, which breaks
the symmetry of the dimer configuration (see e.g. \cite{RMB}). 
Therefore the aim of this paper is to go beyond the symmetric case 
and to connect the properties of the eigenstates of a generalized and 
symmetry-broken spin-boson Hamiltonian with the line spectrum 
of absorption bands of asymmetric excitonic-vibronic 
coupled molecular dimers.

We combine a phase space analysis of the eigenstates, 
which is based on the method of Husimi projections \cite{T},  
with the calculation of the matrix elements of optical transitions.
Husimi projections have repeatedly been used
in order to establish the features of nonintegrability 
of classical systems in the corresponding quantum case 
\cite{MSLN,LB,CCLW}. 
Analyzing the Husimi projections of the eigenstates 
of the spin-boson Hamiltonian in relation to
optical transitions we indicate a bridge 
between the eigenstate analysis of nonintegrable systems 
and observables in an optical experiment. 
In particular, we show how the intensity variations in the 
Husimi projections are related to the spectral randomness
in the fine structure of absorption bands. Such spectral randomness
in the eigenstates of the spin-boson Hamiltonian is also known as 
incipience of quantum chaos \cite{K,CCM,SE96}. 
In this case random features
of the spectrum are just appearing while the system is still far
from displaying universal spectral fluctuations described
by random matrix theory and known for excited states 
of spectra of polyatomic molecules \cite{ZKCPD}. 

A central point in our eigenstate analysis will be to
find out to what extent the adiabatic reference
systems of the spin-boson Hamiltonian are present in its exact 
eigenstates and how the appearance of 
spectral randomness can be interpreted as a mixing of
such reference systems. Following \cite{SE96} we show 
that Bloch projections are a useful quantitative characteristic 
to describe this mixing. Computing the Bloch projections of
the eigenstates, it is possible to distinguish the spectral region 
where the adiabatic branches of the spectrum are still intact 
from the region with a substantial mixing of adiabatic reference 
systems. Furthermore, by performing projections 
of the numerically obtained eigenstates onto the ground state
we obtain the details of the fine structure of the absorption bands 
in the spectral regions as indicated by the Franck-Condon principle. 
We note that the asymmetric case
has some special features, which are not present in the symmetric
situation: In the asymmetric case the final states of an 
optical transition can be located in the higher energetic regions 
of the corresponding adiabatic potentials. Then, as will be shown 
below for two representative cases, the fine structure 
of the spectrum and of the absorption bands can greatly vary 
from regular to irregular arrangements of absorption lines. 

The paper is organized as follows: In Sec.II the model and the
basic equations are presented. In Sec.III the eigenstate analysis
for two representative cases of the asymmetric spin-boson Hamiltonian
is performed and in Sec.IV this analysis is connected with the 
properties of the absorption bands in the corresponding spectral
regions. Finally, in Sec.V our conclusions are summarized.   

\section{Model}

We consider a dimer system described by the Hamiltonian

\begin{equation}
\label{dimer}
\hat{H} = \sum_{n}(\epsilon_n + \gamma_n q_n)|n \rangle \langle n|
+ \sum_{\stackrel{\scriptstyle n,m}{ n \neq m}} V_{n m}|n \rangle \langle m| + 
\sum_{n}\frac{1}{2}(p_n^2 + \omega_n^2 q_n^2),
\end{equation}
where $ |n\rangle $ ($|m\rangle $) are the excited states 
of the molecular monomers constituting the dimer ($n,m = 1,2$) 
with the excitonic energies $\epsilon_n$, 
the coupling constants to intramolecular vibrations $\gamma_n$ 
and  the transfer matrix element $V_{n m}$.
The variables  $q_n$, $p_n$  and $\omega_n$ represent coordinate, 
momentum and frequency of the vibrations, respectively. 
The coupling of the excited dimer states to an incident light 
wave with electrical field vector $\vec{E}_n(t)$ is given by

\begin{equation}
\label{int}
\hat{H}_{int} = \sum_{n}\vec{\mu}_n \vec{E}_n(t)|n \rangle \langle 0|,
\end{equation}

where $\vec{\mu}_n$ and $\vec{E}_n(t)$ 
are the optical transition matrix element 
and electric field at the nth monomer, respectively, and $ |0\rangle $ is
the excitonic vacuum. 
We consider an asymmetric configuration of the dimer system by 
assuming different monomer energies $\epsilon_n$ and optical transition
matrix elements $\vec{\mu}_n$. 
In order to keep the number of asymmetry parameters minimal we
assume symmetric transfer rates, 
$V_{n m} = V_{m n} = -V$, $V > 0$, and 
equal coupling constants and frequencies, 
i.e. $\gamma_1 = \gamma_2  = \gamma$ and $\omega_1=\omega_2=\omega$.
Then the coupling can be reduced to a single vibrational 
mode by introducing the relative displacement of the vibrations. 

The Hamiltonian (\ref{dimer}) can be represented as an operator in the
space of two-dimensional vectors $C={c_1 \choose c_2}$ using the 
standard Pauli spin matrices $\hat{\sigma}_i$ ($i=x,y,z$). 
Passing to dimensionsless variables by 
measuring the energy in units of $2V$, $H = H / 2V$, one obtains 
from (\ref{dimer}) the spin-boson Hamiltonian

\begin{equation} 
\label{Spinboson}
\hat{H} = \epsilon_+\openone  - \frac{1}{2}\hat{\sigma}_x + 
\frac{1}{2}(\hat P^2 + r^2 \hat Q^2)\openone 
 + (\sqrt{\frac{p}{2}}r\hat Q + \epsilon_-)\hat{\sigma}_z .
\end{equation}

In (\ref{Spinboson}), the dimensionless relative displacement and 
corresponding momentum, to which the exciton is coupled,  
are given by $\hat Q = \sqrt{2V} (\hat q_1 -\hat q_2)$
and $\hat P = (\hat p_1 - \hat p_2)/\sqrt{2V}$. 
The dimensionless 
parameters of the spin-boson Hamiltonian (\ref{Spinboson}) 
are related to the dimer system (\ref{dimer}) by 

\begin{eqnarray}
p &=& \frac{\gamma^2}{2V\omega^2}\; ,  \\         
r &=& \frac{\omega}{2V} \;  ,   \\                    
\epsilon_{\pm} &=& \frac{\epsilon_1 \pm \epsilon_2}{4V} \; .
\end{eqnarray}

The parameter $p$ expresses the coupling strength, and $r$ is the
adiabatic parameter measuring the relative strength of quantum effects
of the subsystems. 
The sum $\epsilon_+$ represents the center of 
the excitation energy, whereas the difference $\epsilon_-$ is the 
asymmetry parameter. 
For $\epsilon_-=0$, one obtains the symmetric spin-boson Hamiltonian with 
conservation of total parity (given by the 
operation $\hat Q \rightarrow -\hat Q$,
 $\hat{\sigma}_z \rightarrow -\hat{\sigma}_z$). 
For $\epsilon_- \neq 0$, this symmetry is broken. 

The adiabatic reference systems associated with (\ref{Spinboson})
are introduced by considering the eigenvalue problem of the adiabatic 
part $\hat{h}_{\mathrm{ad}}$, 

\begin{equation}
\label{adpart}
\hat{h}_{\mathrm{ad}} = - \frac{1}{2}\hat{\sigma}_x + 
(\sqrt{\frac{p}{2}}rQ + \epsilon_-)\hat{\sigma}_z 
+  \frac{r^2}{2} Q^2\openone\; .
\end{equation}

The $Q$-dependent eigenvalues and eigenstates of $ \hat{h}_{\mathrm{ad}}$ 
are easily obtained. In particular, one finds the two adiabatic potentials 
 $U^{\pm}_{\mathrm{ad}}(Q)$ for the upper (+) and lower (-) states

\begin{equation}
\label{adpotential}
U^{\pm}_{\mathrm{ad}}(Q) = \frac{r^2}{2}Q^2 \pm 
\sqrt{\frac{1}{4}+\left(\epsilon_- + \sqrt{\frac{p}{2}}rQ \right) ^2} \; ,
\end{equation}

from which the Hamiltonians of the adiabatic reference systems 
are obtained

\begin{equation}
\label{adhamil}
H^{\pm}(Q) = \frac{1}{2}P^2 + U^{\pm}_{\mathrm{ad}}(Q).
\end{equation}

A key point in our analysis of the eigenstates and absorption bands
will be the extent to which the exact eigenstates 
of the Hamiltonian (\ref{Spinboson}) can be viewed as a mixing of adiabatic 
reference systems connected with the adiabatic Hamiltonians (\ref{adhamil}).
As is shown below, the extent of this mixing 
can be controlled by analyzing Husimi projections and  
by computing Bloch projections from the numerically obtained eigenstates.
Husimi projections constitute phase space representations of the eigenstates
which can be compared with the phase space orbits 
of the adiabatic reference systems (\ref{adhamil}). 
Bloch projections constitute another independent indicator of
the mixing of the adiabatic branches in the spectrum. 
For the adiabatic branches these projections are calculated 
from the eigenstates of the Hamiltonian (\ref{adpart}) 

\begin{mathletters}
\label{adstates}
\begin{eqnarray}
|\varphi^+_{\mathrm{ad}}(Q)\rangle &=& \frac{1}{\sqrt{2}} \left(
\sqrt{1+A(Q)}|\uparrow \rangle - \sqrt{1-A(Q)}|\downarrow \rangle \right)
  \;\label{adstate+} \\
|\varphi^-_{\mathrm{ad}}(Q)\rangle &=& \frac{1}{\sqrt{2}} \left(
\sqrt{1-A(Q)}|\uparrow \rangle + \sqrt{1+A(Q)}|\downarrow \rangle \right)
  \; \label{adstate-}
\end{eqnarray}
\end{mathletters}

where

\begin{equation}
\label{adstatecoeff}
A(Q)=\frac{2\epsilon_- + \sqrt{2p}rQ}
{\sqrt{1+(2\epsilon_- + \sqrt{2p}rQ)^2}}.
\end{equation}

Computing the expectation values 
$x = \langle \hat{\sigma}_x \rangle$ 
of the Pauli matrix $\hat{\sigma}_x$ 
for the adiabatic eigenstates (\ref{adstates}), one 
obtains the Bloch projections of the adiabatic branches

\begin{equation}
\label{xad}
\langle \varphi^{\pm}_{\mathrm{ad}}(Q) | \hat{\sigma}_x | 
\varphi^{\pm}_{\mathrm{ad}}(Q)\rangle = \mp \frac{1}
{\sqrt{1+(2\epsilon_-+\sqrt{2p}rQ)^2}} \; ,
\end{equation}

According to eq.(\ref{xad}) there is a distinction 
in the sign of the Bloch projections for the adiabatic eigenstates. 
Computing the Bloch projections from the numerically obtained
eigenstates this distinction in the sign will be used
as an indicator for the case in which the spectrum resolves 
into adiabatic branches. 

\section{Eigenstate analysis}

Our analysis is based on a diagonalization of the spin-boson Hamiltonian 
using a basis of product states of spin states with harmonic oscillator 
eigenstates. 
In this basis the eigenstates are obtained in the form
\begin{equation}
\label{eigstate}
\left| \lambda \right\rangle = \sum_m \left(c_{\uparrow m}^{(\lambda)}
\left| \uparrow \right\rangle + c_{\downarrow m}^{(\lambda)}
\left| \downarrow \right\rangle \right) \left| m \right\rangle \ 
= \sum_m \left|s_{m} ^{(\lambda)}\right\rangle \left| m \right\rangle \ ,
\end{equation}
where $\left| \uparrow \right\rangle  =  {1 \choose 0}$,
 $\left(\left| \downarrow \right\rangle  = {0 \choose 1}\right)$ 
denote the spin up 
(down) states, $\left| m \right\rangle$ the harmonic oscillator eigenstates
and $ c_{z m}^{(\lambda)}$ ($z=\uparrow,\downarrow $) 
the expansion coefficients of the $\lambda$ eigenstate. 
For a fixed $m$ we will also use the spin representation   
$\left|s_{m}^{(\lambda)}\right\rangle$ for the up and down coefficients 
$c_{z m}^{(\lambda)}$ ($z=\uparrow,\downarrow$) for a given 
oscillator state as indicated in the second part of   
(\ref{eigstate}). The matrix dimension in the numerical 
diagonalization was $N=4000$, for the eigenstate analysis
the first $1100$ eigenstates were used. 

For the phase space representation of the eigenstates 
Husimi projections are used. Husimi projections 
$h_{\lambda}\left(\alpha(Q,P);s\right)$ 
are defined by projecting an eigenstate $|\lambda\rangle$ 
onto a set of coherent oscillator states $|\alpha(Q,P)\rangle$,
which scan the oscillator phase plane of the vibrational 
subsystem, while the spin projection 
 $|s\rangle = c_{\uparrow}\left|\uparrow\right\rangle +  c_{\downarrow}\left|\downarrow\right\rangle$ 
is kept fixed, 
\begin{equation}
\label{husia}
h_{\lambda}(\alpha(Q,P);s) = 
| \langle \, \lambda \: |\: \alpha(Q,P), s \rangle|^2,
\end{equation}
and  $|\alpha(Q,P),s \rangle =|\alpha(Q,P) \rangle |s\rangle$.
Using the explicit form of $\left|\lambda\right\rangle$ in 
(\ref{eigstate}) one finds 
\begin{equation}
\label{husimi}
h_{\lambda}(\alpha(Q,P);s) = \left|
\sum_m (c_{\uparrow m}^{\ast (\lambda)} c_{\uparrow} +
c_{\downarrow m}^{\ast (\lambda)} c_{\downarrow}) 
\left\langle m | \alpha(Q,P) \right\rangle \right|^2,
\end{equation}
where $\left\langle m | \alpha(Q,P) \right\rangle 
= \frac{\alpha^m}{\sqrt{m{\mathrm !}}}{\mathrm e}^{-\frac{|\alpha|^2}{2}}$ 
and $\alpha(Q,P)=\sqrt{\frac{r}{2}}\langle \hat Q \rangle 
+ \frac{i}{\sqrt{2r}}\langle \hat P \rangle$.

For the Bloch projection of an eigenstate, 

\begin{equation}
\label{xlambda}
x_{\lambda} = \left\langle\lambda \right|\hat{\sigma}_x \left|\lambda\right\rangle,
\end{equation}

one finds from the expansion (\ref{eigstate}) 

\begin{equation}
\label{xbloch}
x_{\lambda}=\sum_m \left( c_{\uparrow m}^{\ast (\lambda)} 
c_{\downarrow m}^{(\lambda)} + c_{\downarrow m}^{\ast (\lambda)}
 c_{\uparrow m}^{(\lambda)} \right)\; .
\end{equation}

Before turning to the eigenstate analysis for definite parameter sets
we note that the appearance of spectral irregularities
can be expected in the 
high energetic region of the overlap of both adiabatic potentials only, 
where a sufficient number of states of both potentials mix. It is 
obvious that such a region can be probed by the final states of 
an optical transition in the asymmetric case only: In the 
symmetric case, as is evident from 
the adiabatic potentials (\ref{adpotential}), 
the optical transitions would necessarily terminate in the ground 
state region of the upper potential, where only a few states of the 
upper potential can mix with the lower potential. 
Therefore when investigating 
the structure of the spectrum and eigenstates of the 
spin-boson Hamiltonian we paid special attention to a systematic 
change of the asymmetry parameter $\epsilon_-$  
combined with the coupling 
parameter $p$ in order to produce appropriate configurations of the 
adiabatic potentials. Such configurations were found for 
relatively high coupling $p$ and asymmetry values $\epsilon_-$. 
Below we present the eigenstate analysis for two 
typical asymmetric configurations in the adiabatic 
parameter region ($r<1$) with the parameter sets 

A: $p=4$, $r=0.1$ and $\epsilon_- =5$ and  

B: $p=20$, $r=0.1$ and $\epsilon_- =10$. 

\begin{figure}
\centerline{\epsfig{file=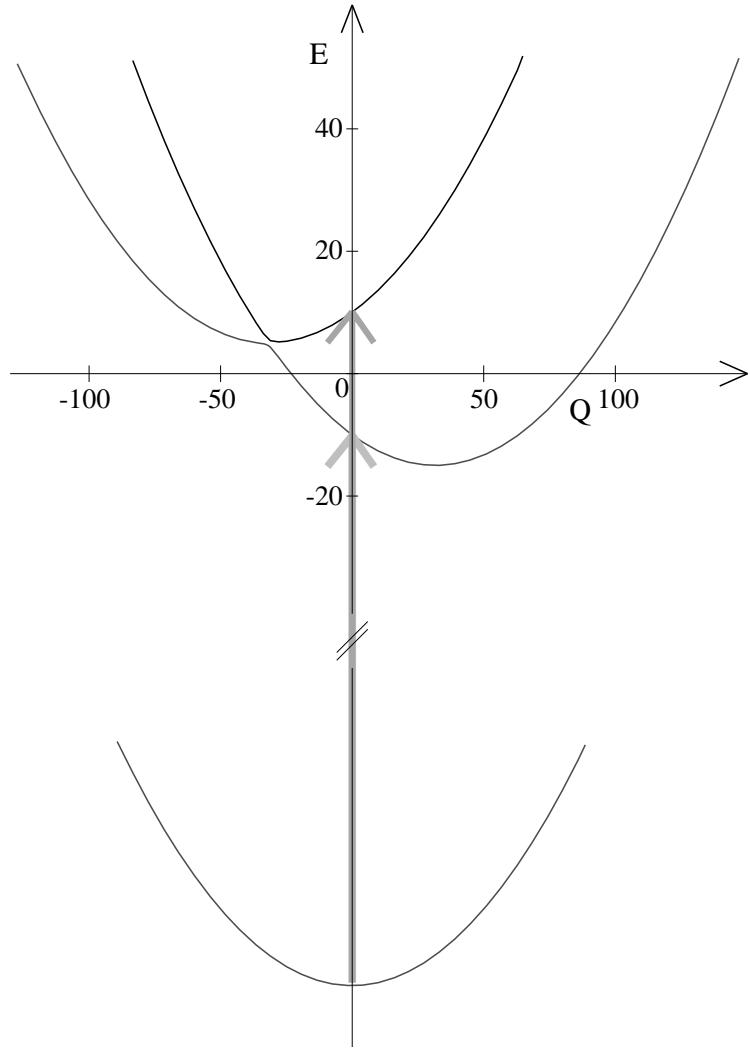,width=10cm}}
\caption{
Asymmetric adiabatic potentials and ground state 
configuration for the parameter set B. In the 
case of the parameter set A the asymmetry in the
configuration of the adiabatic potentials is similar,
but less pronounced as compared to B. 
Arrows mark the location of the absorption bands at
the Franck-Condon energies.}
\label{figadpot}
\end{figure}

The parameter set B corresponds to a stronger coupling value $p$
and a larger asymmetry parameter $\epsilon_-$, as compared to set A.
To indicate the location of the absorption bands, which will be 
analyzed in section~\ref{secabsorption}, 
the adiabatic potentials of set B and the 
ground state of the system are shown in Fig.~\ref{figadpot}.

\begin{figure}
\centerline{\epsfig{file=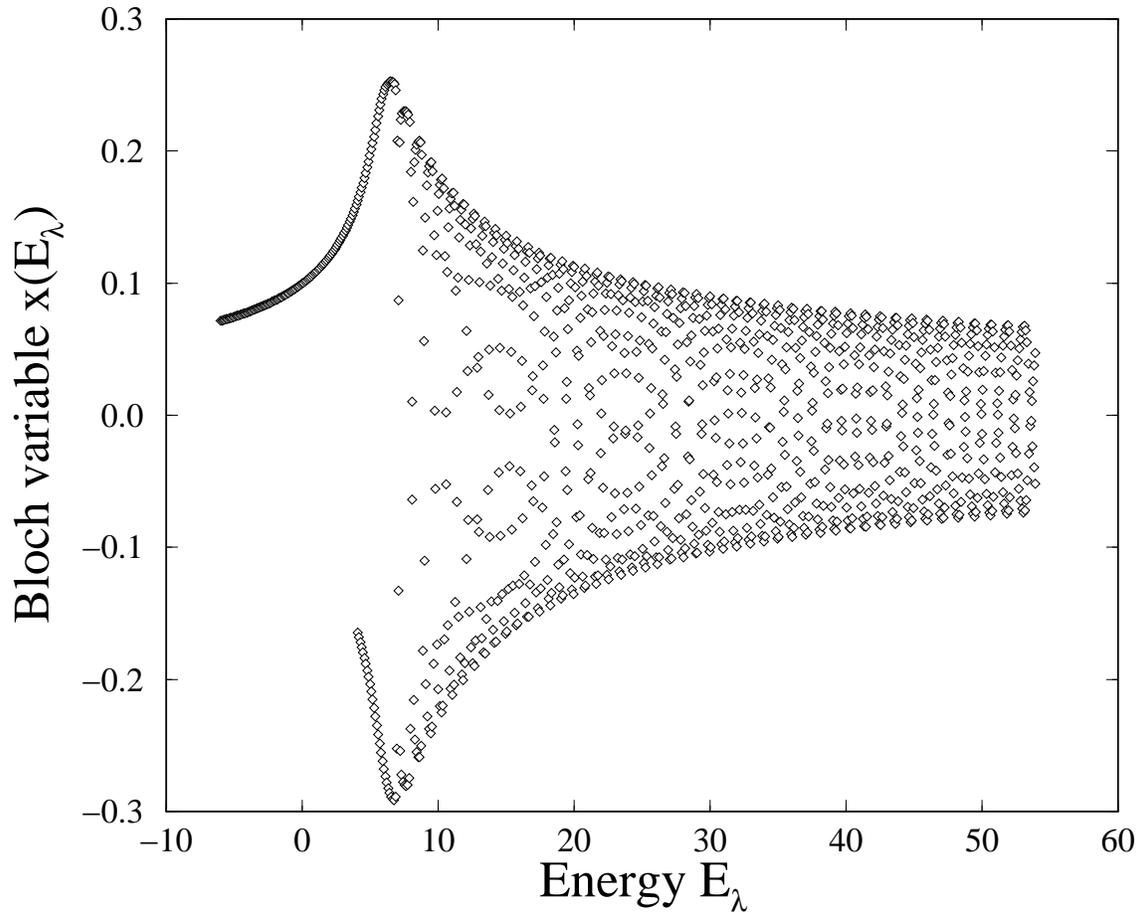,width=15cm}}
\caption{
Bloch projections $x(E_{\lambda})$ of the eigenstates
for the parameter set A. The presence of three
characteristic regions with one adiabatic branch,
coexistence of two adiabatic branches and the 
mixing region is clearly visible. The Bloch
projections for the parameter set B show a similar
behavior.}
\label{figblochA}
\end{figure}

\begin{figure}
\centerline{\epsfig{file=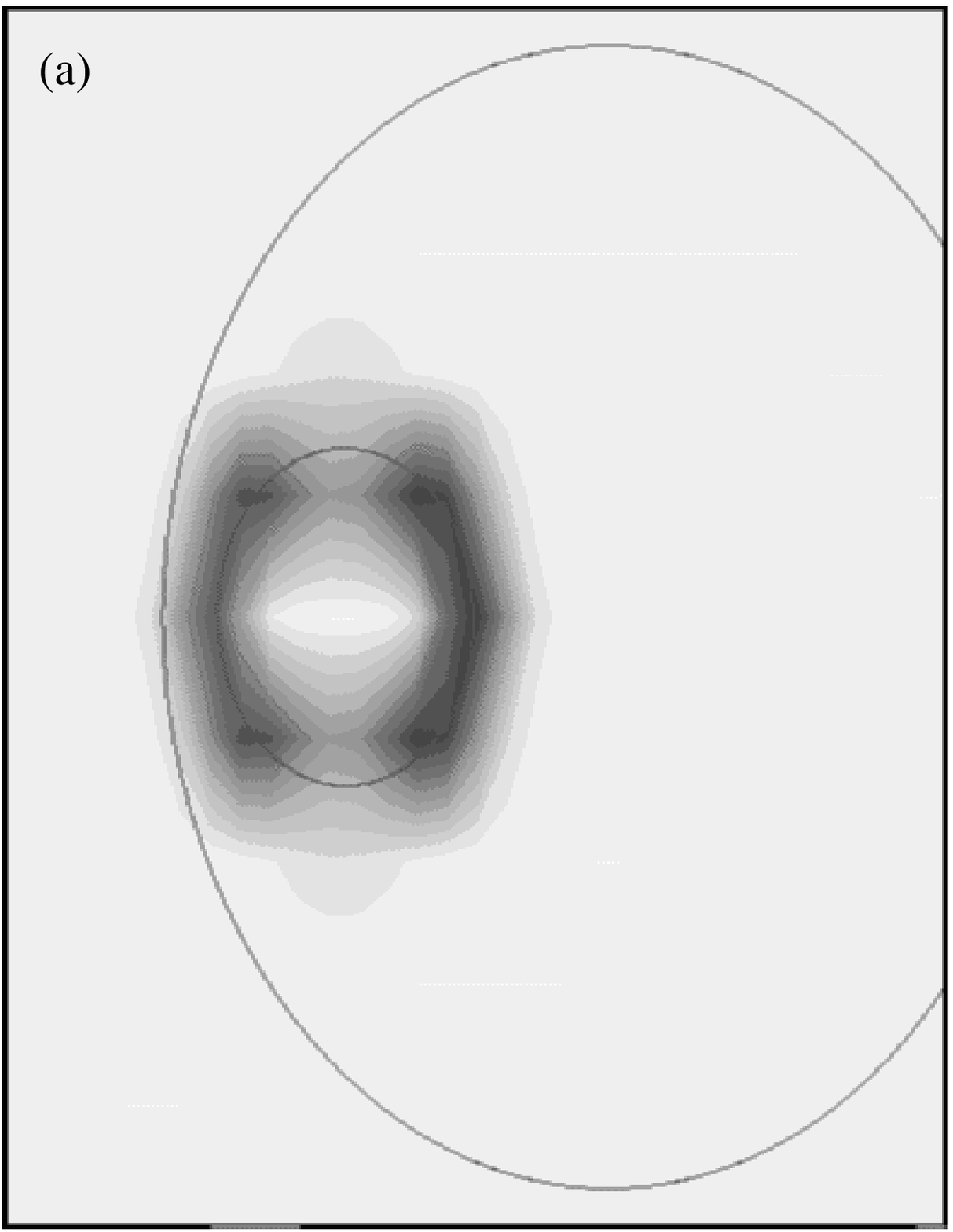,width=12cm,height=6.7cm}}
\centerline{\epsfig{file=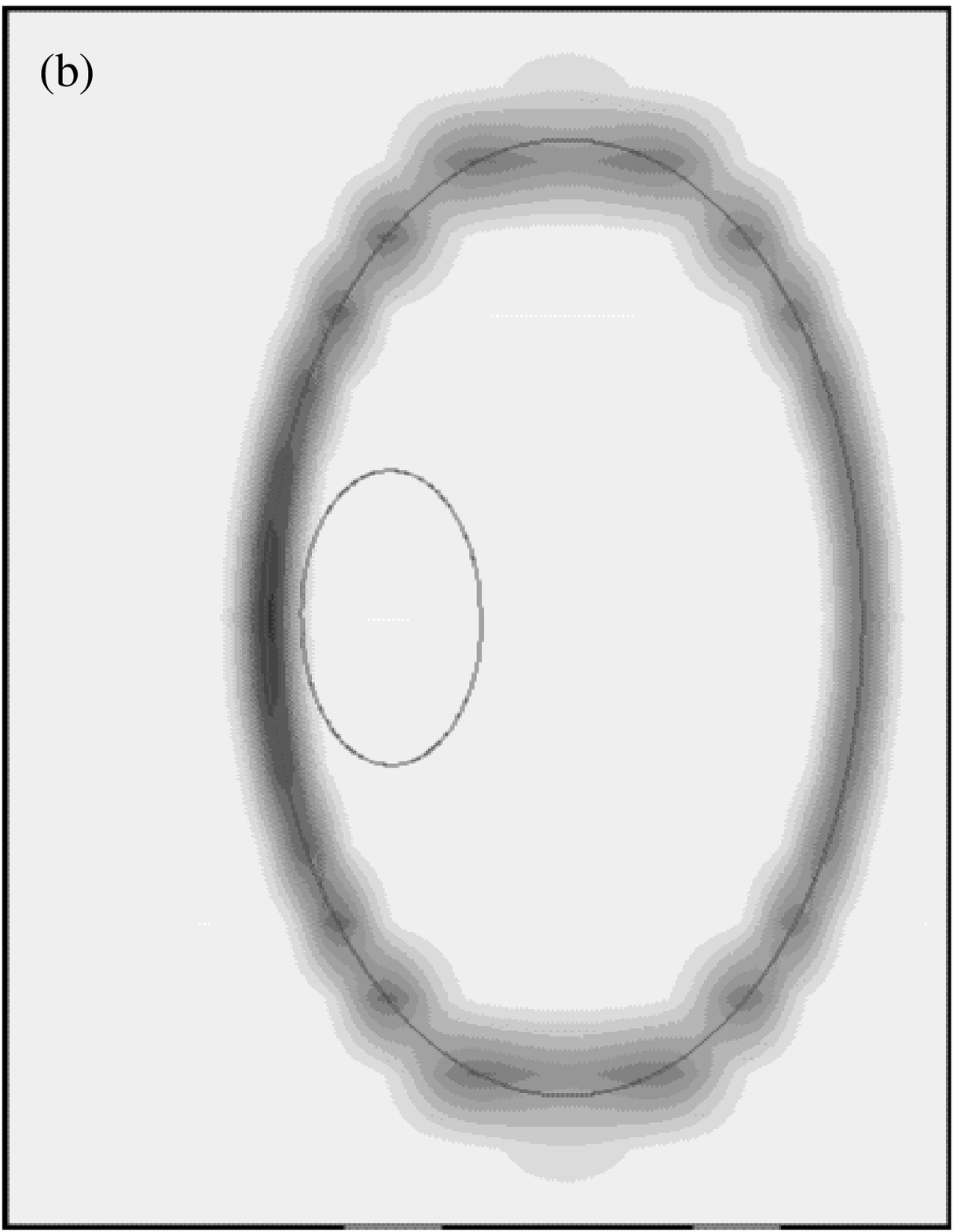,width=12cm,height=6.7cm}}
\centerline{\epsfig{file=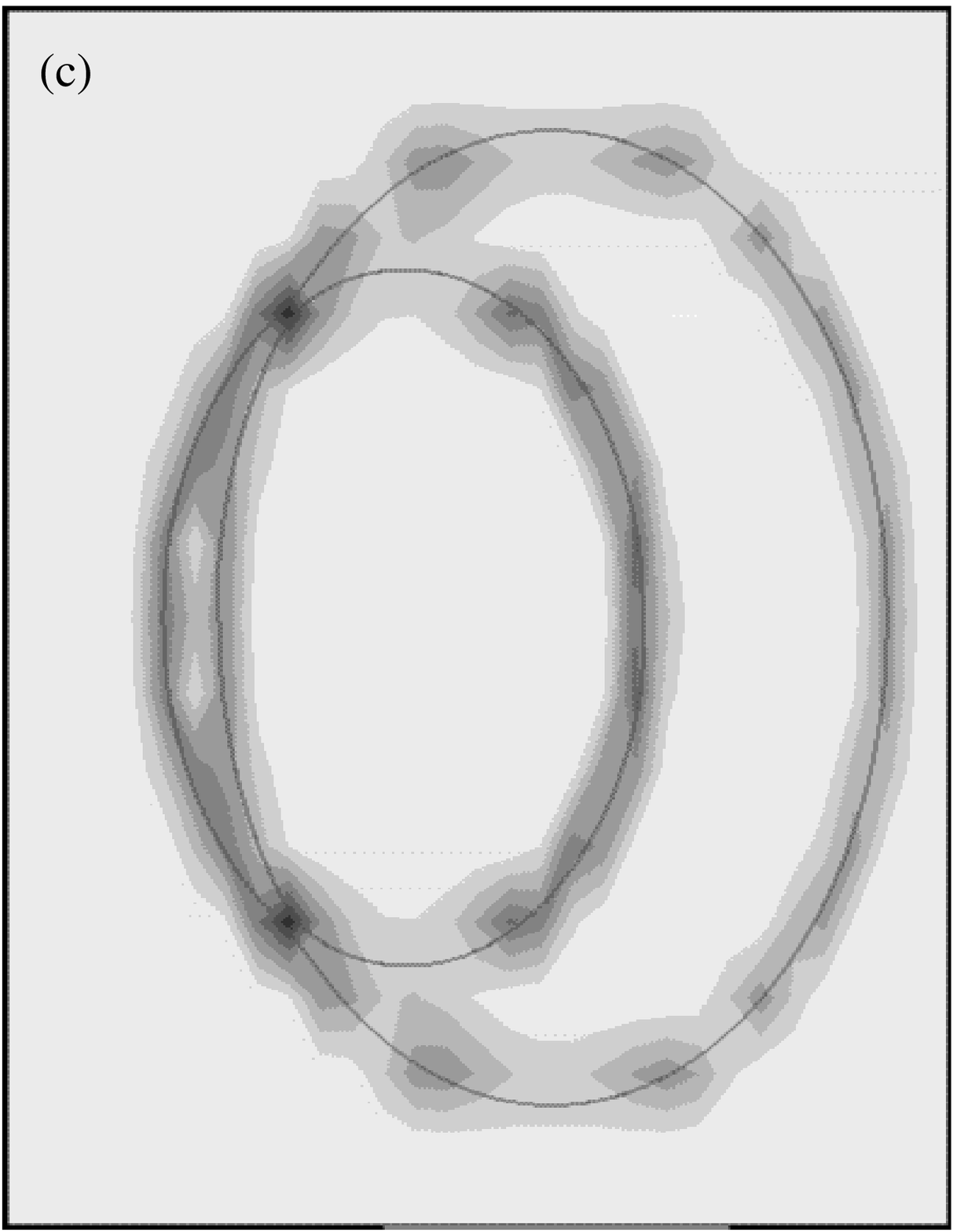,width=12cm,height=6.7cm}}
\caption{
Husimi projections for the parameter set A 
in the overlap region 
of the adiabatic potentials. The solid lines indicate the
classical phase space orbits of the
adiabatic Hamiltonians (\ref{adhamil}).
The projections correspond to the eigenstates
120 (a), 121 (b) and 310 (c),
with the eigenstate energies 
$E_{120}=5.0028$, $E_{121}=5.0956$ and $E_{310}=14.5176$,
respectively.   
The selected phase space parts are
 (a): $-50\le Q\le50$, $-5\le P\le5$,
 (b): $-75\le Q\le75$, $-6\le P\le6$ and
 (c): $-90\le Q\le90$, $-8\le P\le8$,
with the Q-axis displayed horizontally and P-axis 
displayed vertically. 
The projections in (a) and (b) are located 
on the phase space orbits of the upper and lower 
adiabatic potentials, respectively, and show
the coexistence of both adiabatic branches in 
the selected energy interval. The projection (c)
is located on the phase space orbits of 
both adiabatic potentials and characteristic
for the mixing region of the adiabatic 
reference states in the spectrum.}
\label{fighus_A}
\end{figure}

In Fig.~\ref{figblochA} we start with the eigenstate analysis for the 
parameter set A by displaying the Bloch projections $x(E_{\lambda})$
computed from the eigenstates according to eq.(\ref{xbloch}).
In the low energy region one finds one adiabatic branch 
in the $x(E_{\lambda})$ dependence associated with the 
lower adiabatic potential (for which $x_{\lambda}>0$ in accordance 
with the sign in (\ref{xad})). The overlap of the two potentials 
is marked by the appearance of
a second adiabatic branch in the $x(E_{\lambda})$ dependence.
The mixing of the eigenstates of the two adiabatic potentials
is seen by the disappearance of both adiabatic branches and 
indicated by a broad band of Bloch projections in the high energy region.
The presence of three characteristic regions in the Bloch projections
with one adiabatic branch, two adiabatic branches and the mixing region
is reflected in the Husimi projections of the eigenstates. In the 
regions where the adiabatic branches are intact one obtains regular
Husimi projections concentrated around the phase space orbits
of the integrable adiabatic reference Hamiltonians (\ref{adhamil}).
Typical examples of such regular projections corresponding 
to equal spin projections $c_{\uparrow}=c_{\downarrow}$
for the energetic overlap region with two adiabatic branches present 
are shown 
in Fig.~\ref{fighus_A}(a) and Fig.~\ref{fighus_A}(b). 
In Fig.~\ref{fighus_A}(a) 
the Husimi projection corresponding to an eigenstate in 
the upper adiabatic potential is shown, whereas 
in Fig.~\ref{fighus_A}(b)
a projection associated with an eigenstates of the lower 
adiabatic potential is displayed. 
Sequences of such projections clearly attributable to the 
coexistence of two independent adiabatic branches were observed until  
the mixing region is reached. Then Husimi projections incorporating
the phase space features of both adiabatic potentials are
observed, as shown in Fig.~\ref{fighus_A}(c). 

A closer inspection of these projections in the mixing region
shows that although the distribution is concentrated on the
regular phase space orbits of the adiabatic potentials, 
the intensity between these parts varies in an irregular 
and random way, when Husimi projections
for a sequence of eigenstates are considered.
The structure of the eigenstates which shows up 
in the random shift of intensity in the Husimi projections 
is a source of spectral randomness in the absorption bands. 
In the case of the parameter set A, however,
this region of an irregular and random shift of the phase space distribution
is not reached by the absorption processes. In this case the absorption band 
is located in the region, where the two adiabatic branches are still intact 
and no spectral mixing between their eigenstates occurs.   

\begin{figure}
\centerline{\epsfig{file=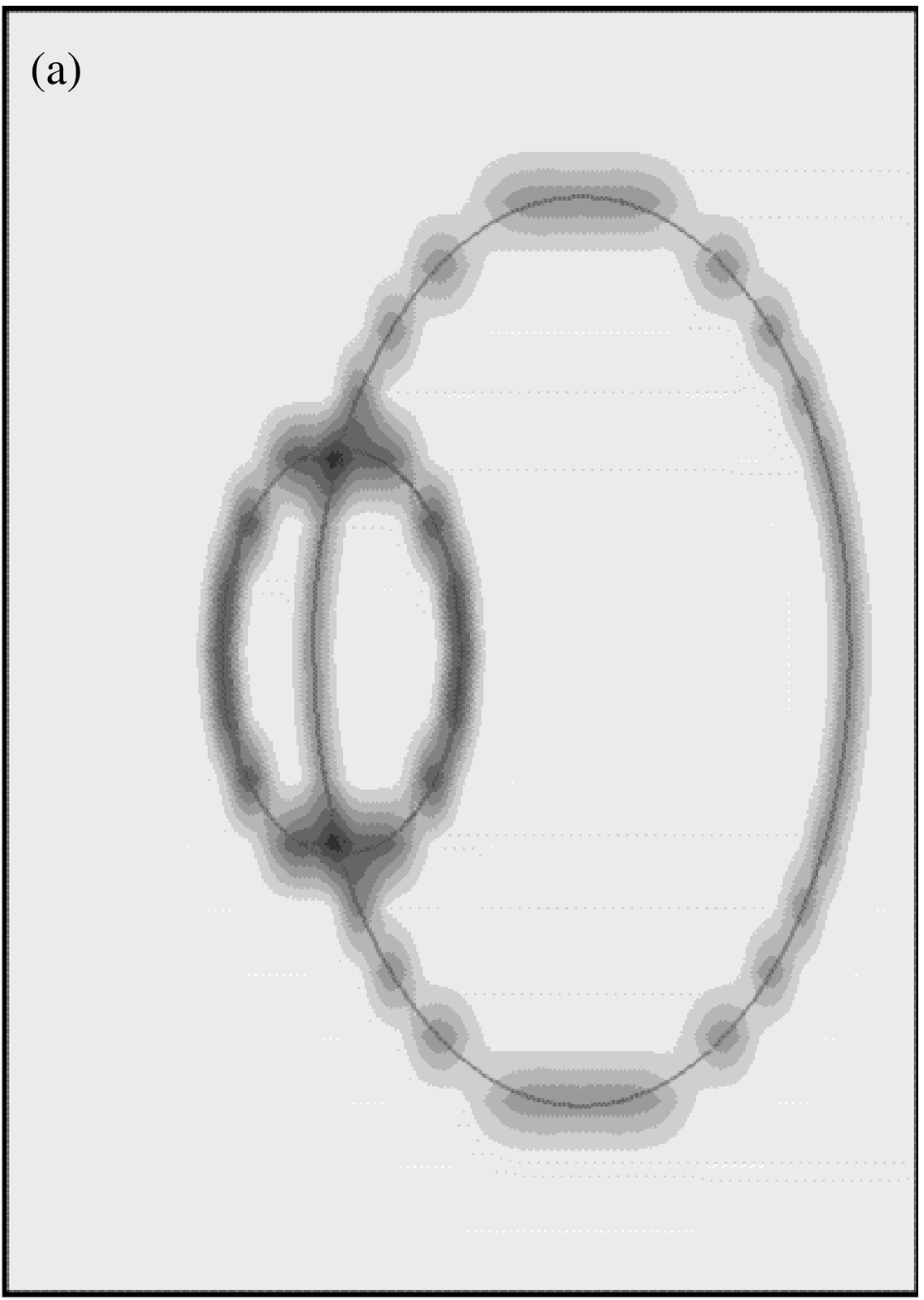,width=12cm,height=6.6cm}}
\centerline{\epsfig{file=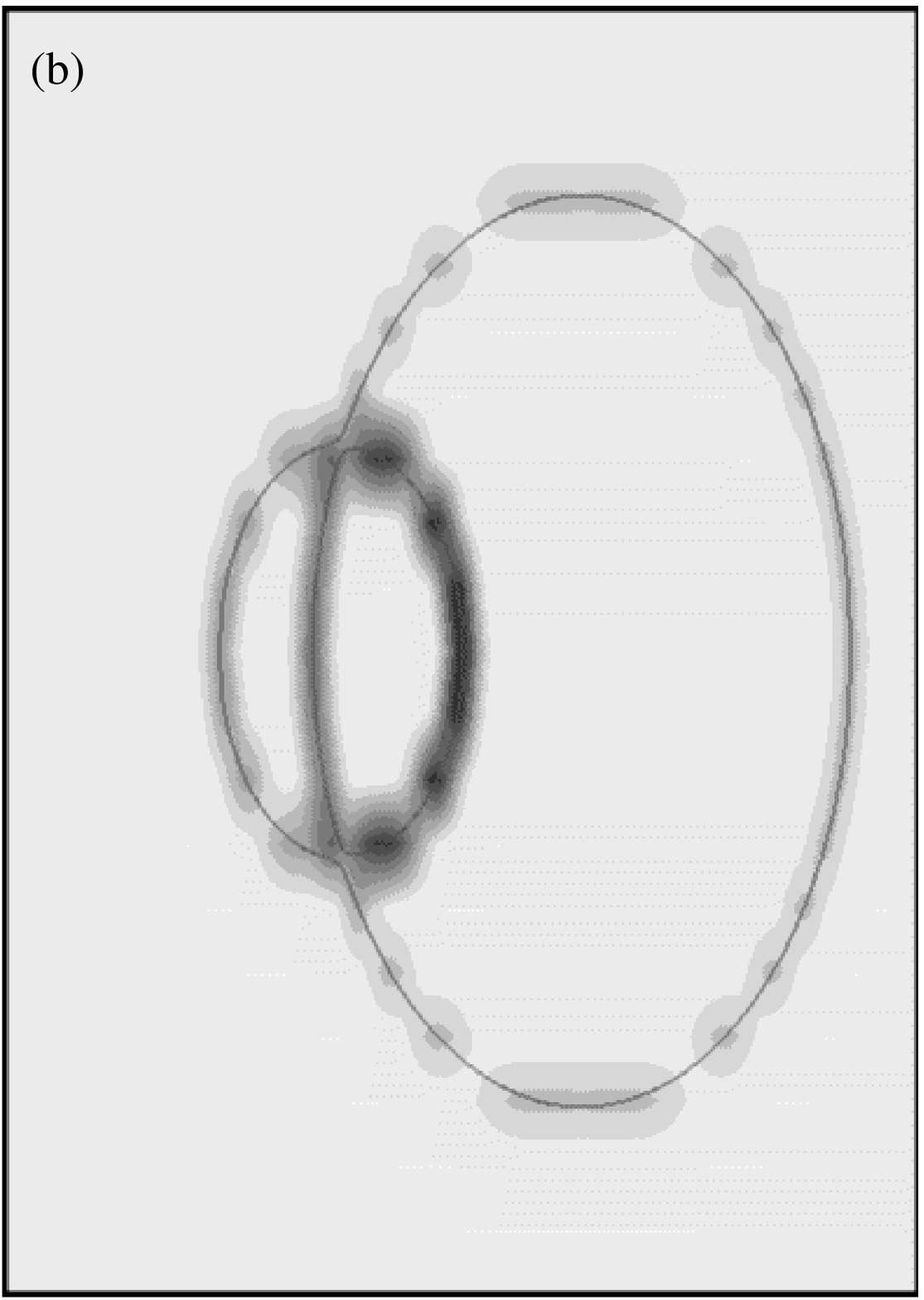,width=12cm,height=6.6cm}}
\centerline{\epsfig{file=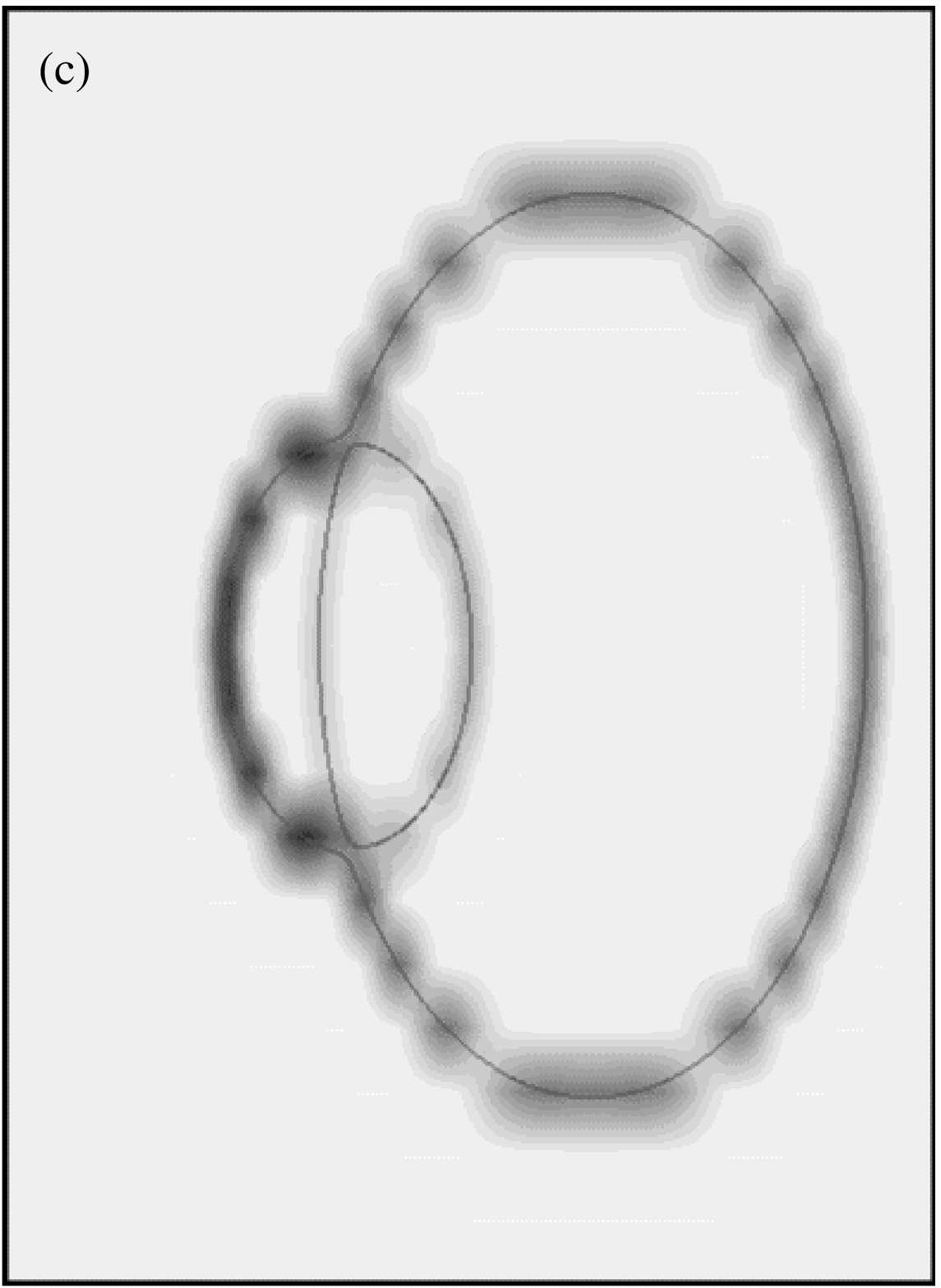,width=12cm,height=6.6cm}}
\caption{
Husimi projections for the parameter set B.
The solid lines indicate the classical phase space orbits 
of the adiabatic Hamiltonians (\ref{adhamil}). 
The projections correspond to the
eigenstates 300 (a), 302 (b) and 303 (c), 
with the eigenstate energies 
$E_{300}=10.0237$, $E_{302}=10.1467$ and $E_{303}=10.1520$,
respectively.   
In all projections the selected phase space parts 
are $-120\le Q\le120$, $-10\le P\le10$,
with the Q-axis displayed horizontally and P-axis 
displayed vertically. 
The projections are located on the phase space orbits
of both adiabatic potentials and characteristic for
the mixing of adiabatic reference states. One observes 
a random variation in the intensity of the distribution 
between the different eigenstates.
This random variation includes the center of the rectangle,
i.e. the region around $Q=P=0$, which is the final state
region for the upper absorption band shown in Fig.\ref{figabs_B1}.
The different intensities in this region correspond to
different strengths of the absorption lines of the upper band 
in Fig.\ref{figabs_B1}.}
\label{fighus_B}
\end{figure}

In the case of the parameter set B one finds a situation, 
where the spectral window, which is cut by the transition matrix 
elements of the absorption process, reaches the region of an 
irregular mixing of adiabatic states. 
The Husimi projections for selected eigenstates 
and equal spin projections which correspond to such 
a region and which are relevant as final states of an optical transition 
are shown in the set of Fig.~\ref{fighus_B}(a)-(c). 
One observes a random variation of the intensity in the 
Husimi distributions of the eigenstates between the 
orbits of the two adiabatic potentials.
As we show in the next section, as a consequence of this 
random variation intensities of lines for optical transitions 
terminating in these states vary randomly.

\section{Absorption}
\label{secabsorption}

In the calculation of the optical absorption 
the dimer system is assumed
to be small compared to the optical wavelength. 
Then the electric field vector is approximately equal 
at the monomer sites of the 
dimer configuration, i.e. $\vec{E}_n(t)=\vec{E}(t)$. 
Introducing the projections $\mu_n$ of the monomer 
transition matrix elements onto the direction of the field polarization, 
i.e. $\vec{E}(t)\vec{\mu}_n = E(t)\mu_n$, ($n=1,2$), 
the interaction of the dimer with the incident light wave is 
represented in the form 

\begin{equation}
\label{dipappr}
\hat{H}_{int}=E(t)\hat{M}_{int},
\end{equation}

where $\hat{M}_{int}$ is the interaction operator

\begin{equation}
\label{Mdef}
\hat{M}_{int}=\sum_{n}|n \rangle \mu_n \langle 0|.
\end{equation}

In the ground state there is no excitonic-vibronic interaction, 
i.e. the ground state of the system 
$|g\rangle$ is given by the direct product of the 
excitonic vacuum $|0\rangle=|exc0\rangle$ and the vibrational ground state. For the 
ground state of the vibrational subsystem we assume the 
zero temperature case, 
where it is in its lowest state  $|m=0\rangle$, with $|m=0\rangle$ 
the $m=0$ Hermitian polynomial of the undisplaced $Q$ oscillator. 
Then $|g\rangle=|exc0\rangle|m=0\rangle$. 
For the optical transition matrix element 
$M_{\lambda g}  = \langle \lambda |\hat{M}_{int}| g \rangle$ 
of the interaction operator (\ref{Mdef}), 
which corresponds to an absorption process from the 
ground state $g$ into an excited state $\lambda$, one then obtains 
using eq.(\ref{eigstate}) the expression 

\begin{equation}
\label{Melem}
M_{\lambda g}= \mu_{1} c_{\uparrow 0}^{(\lambda)} + \mu_{2} c_{\downarrow 0}^{(\lambda)} .
\end{equation}

The matrix element (\ref{Melem}) is representable in the form of a 
particular projection of the spin vector associated with the 
$m=0$ vibrational state in (\ref{eigstate}), 
$\left|s_{0} ^{(\lambda)}\right\rangle$. 
Introducing the angles 

\begin{equation}
\label{cossin}
\cos\alpha=\frac{\mu_1}{\sqrt{\mu_1^2 + \mu_2^2}} ,  
\sin\alpha=\frac{\mu_2}{\sqrt{\mu_1^2 + \mu_2^2}}
\end{equation} 

and a spin vector $|s_{\mu}\rangle$ defined by 

\begin{equation}
\label{smu}
| s_{\mu} \rangle ={\cos\alpha \choose \sin\alpha},
\end{equation}
 
one represents (\ref{Melem}) in the form
\begin{equation}
\label{Melempro}
M_{\lambda g }
= \sqrt{\mu_1^2 + \mu_2^2}  \langle s_{\mu}|s_{0}^{(\lambda)}\rangle.
\end{equation}

Measuring the square of (\ref{Melempro}), 
i.e. the absorption strength of an 
optical transition from $|g\rangle$ to $|\lambda\rangle$, 
in units of $(\mu_1^2 + \mu_2^2)$ one finds for the 
dimensionsless absorption strength $Q_{\lambda g}$

\begin{equation}
\label{abs}
Q_{\lambda g} = \frac {M_{\lambda g}^2}{\mu_1^2 + \mu_2^2}
=(\langle s_{\mu}|s_{0}^{(\lambda)}\rangle)^2.
\end{equation}

Comparing eq.(\ref{abs}) with the expression for the Husimi projection 
(\ref{husimi}) it follows that $Q_{\lambda g}$ is equal to 
a Husimi projection of the $\lambda$-eigenstate 
of (\ref{Spinboson}) taken at the phase 
space point $Q=P=0$ with the spin projection being fixed at $s_{\mu}$ : 
For $Q=P=0$ one obtains $\alpha(Q,P)=0$, 
which selects the $m=0$ term in the sum of (\ref{husimi}), inserting 
the value $s_{\mu}$ for the spin vector $s$, one finds 
that $h_{\lambda}(0;s)$  reduces to the r.h.s. of eq.(\ref{abs}), i.e.

\begin{equation}
\label{abshus}
Q_{\lambda g} = h_{\lambda}(0;s_{\mu}).
\end{equation}

Eq. (\ref{abshus}) is in line with the Franck-Condon principle for 
optical transitions: The transition is vertical from the ground state 
region at $Q=P=0$ in the phase space representation and probes the 
intensity of the final state by its Husimi distribution
in the same region. 
The optical transition matrix elements of the monomers define the
spin projection $s_{\mu}$. For the special case of a symmetric dimer with 
$\mu_1=\mu_2$ the transition occurs into the symmetric combination 
of monomer states.
Differences in the optical transition matrix elements, i.e. 
$\mu_1\not=\mu_2$, introduce a second asymmetry parameter besides that of 
the site energy asymmetry $\epsilon_-$. 
These differences in the optical transition matrix elements express
the optical asymmetry of the dimer and enter the optical dimer matrix
element (\ref{Melem}) through the spin projections (\ref{cossin}).

\begin{figure}
\centerline{\epsfig{file=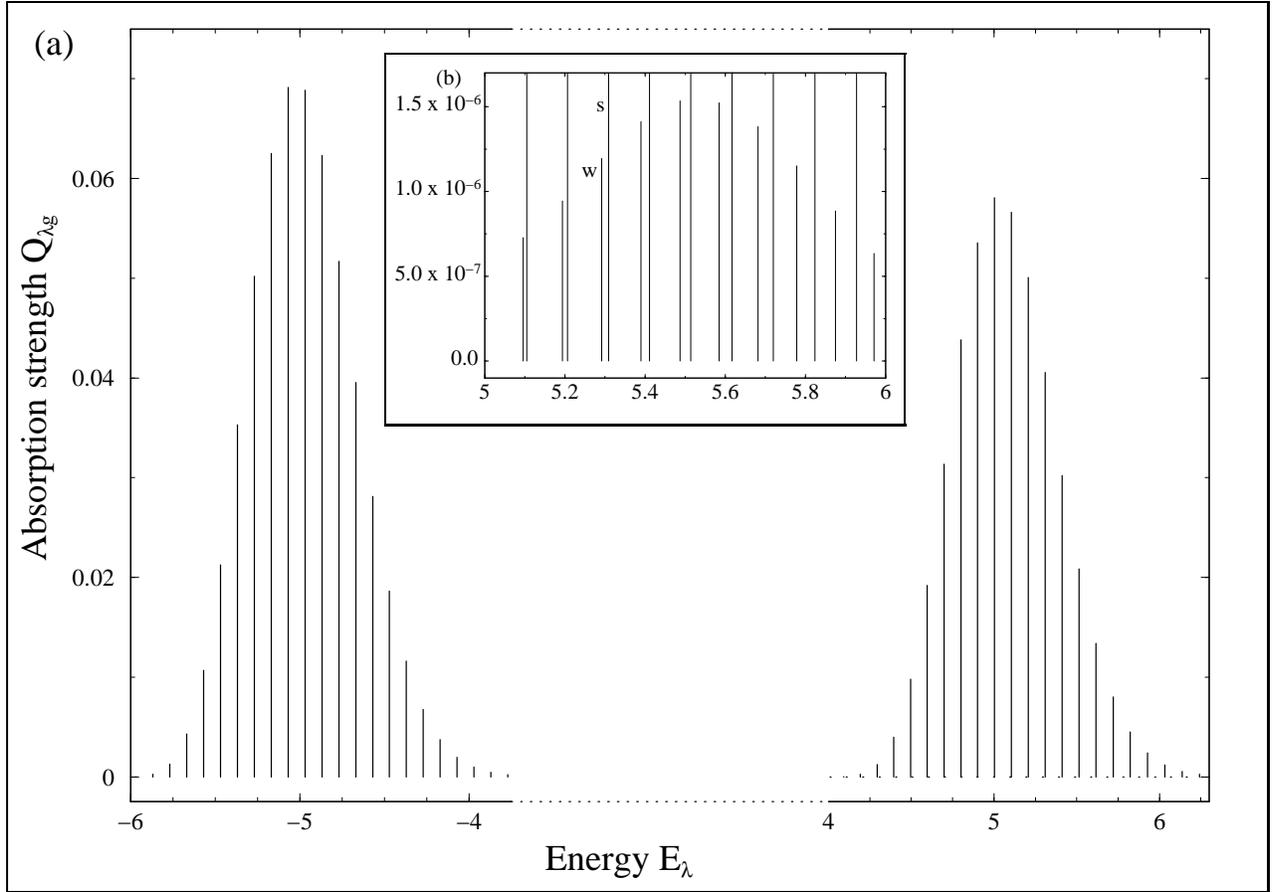,width=12cm,angle=270}}
\caption{
Lower and upper absorption bands for the parameter set A. 
In (a) the strong intensity lines of the upper band are visible
only. A closer inspection of the upper band shows that this band
is a superposition of two regular bands corresponding to the 
final states in the upper and lower potential and the 
coexistence of two independent adiabatic branches 
in this part of the spectrum.
This superposition becomes evident from the inset (b), 
in which a change of scale is used to display the weak lines,
which are embedded between the strong lines (not drawn to
peak intensity and cut off at the upper edge). In the inset
a pair of neighbouring weak and strong lines are indicated by 
the labels w and s, respectively.}
\label{figabs_A}
\end{figure}

In Fig.~\ref{figabs_A}(a) the absorption bands $Q_{\lambda g}$ 
calculated from the numerically obtained eigenstates 
for the parameter set A are represented as stick spectra. 
Absorption bands are located in accordance with the
Franck-Condon principle. 
This is seen by comparing their positions with the final states 
following from the adiabatic potentials in (\ref{adpotential})
by setting $Q=0$. 
One finds an energetically lower band 
for optical transitions terminating in the lower adiabatic potential
and a higher band for transitions occuring into the region of the overlap
of the two potentials. 
The shape of the lower band on the left side of Fig.~\ref{figabs_A}(a)
is completely regular. The lines of the higher band on the right side
of Fig.~\ref{figabs_A}(a) are also regularly arranged. A closer
inspection of the higher band, however, reveals that between the lines
shown in Fig.~\ref{figabs_A}(a) lines with a much weaker intensity 
are embedded, which form another regular band. This is evident from the
inset, Fig.~\ref{figabs_A}(b), where parts of the lines of the weak
and strong bands are shown together. In order to make the weak band visible  
in the inset a much smaller intensity scale is used. 
The two bands, into which the absorption spectrum 
of the high energy part resolves, are easily identified 
as to belong to the states of the two coexisting adiabatic branches 
analyzed in the preceding section for the parameter set A: 
The weak band with relatively small intensities is due to 
optical transitions terminating in the 
high energy states of the lower 
potential (the sign of the Bloch variables of the
final states is $x_\lambda>0$ in accordance with the sign 
of the lower adiabatic branch (\ref{xad})), 
whereas the strong band with much greater intensities is due to
optical transitions into the low energy states of the 
upper potential (for these states $x_\lambda<0$,
again in accordance with the sign of the adiabatic branch (\ref{xad})).

\begin{figure}
\centerline{\epsfig{file=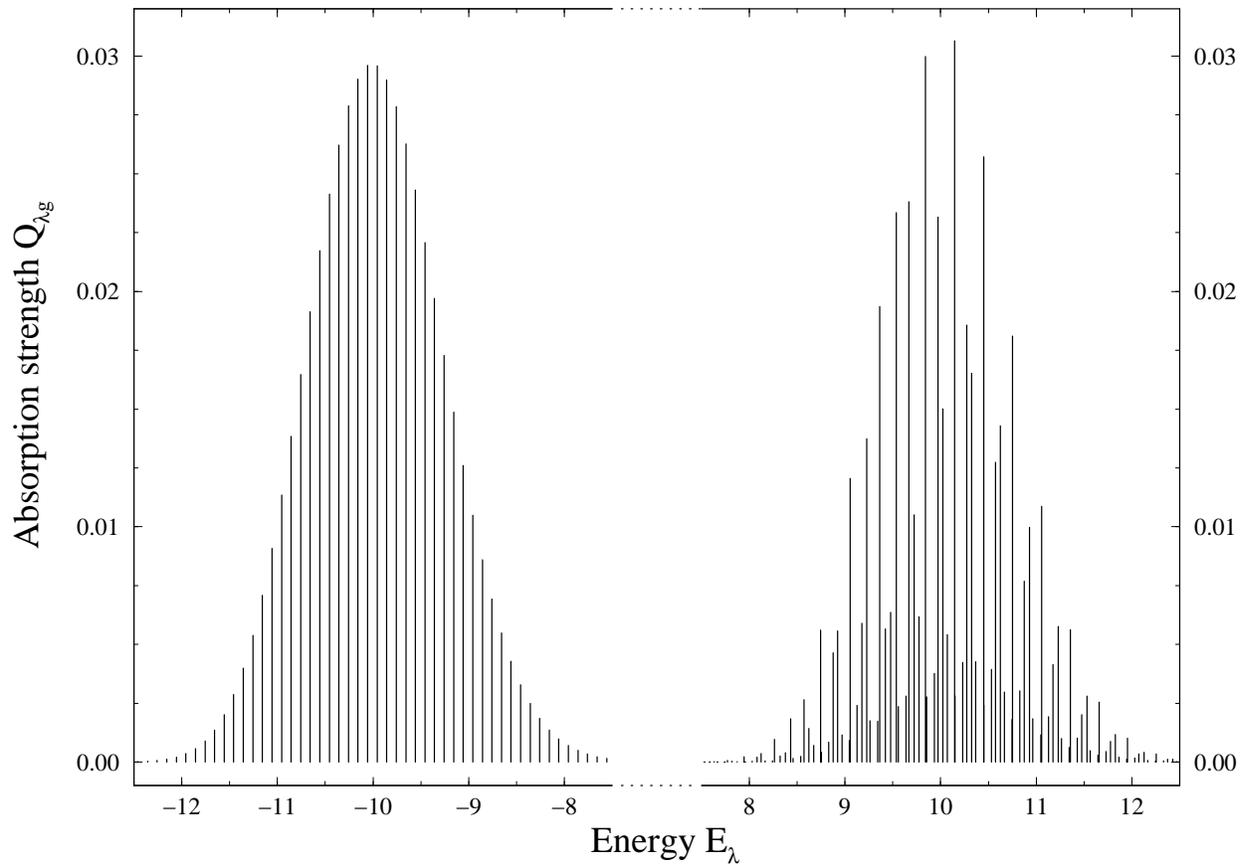,width=12cm,angle=270}}
\caption{
Lower and upper absorption bands for the 
parameter set B and the case of optical
symmetry, $\mu_1=\mu_2$. A broken energy scale 
is used to display both bands. 
The lower band is regular with
final states in the lower adiabatic potential,
whereas the upper band is irregular due to the
mixing of the adiabatic reference states, compare 
with the intensity variation in the Husimi 
projections displayed in Fig.~\ref{fighus_B}.}  
\label{figabs_B1}
\end{figure}

In Fig.~\ref{figabs_B1} the absorption bands are shown 
for the parameter set B. The bands are located in the 
spectral regions of the Franck-Condon energies indicated
by the arrows in Fig.~\ref{figadpot}.
The lower absorption band, 
shown on the left side of the Fig.~\ref{figabs_B1}, 
is regular like the lower band in the Fig.~\ref{figabs_A}(a). 
The higher band,
displayed on the right side of Fig.~\ref{figabs_B1}, however,  
is completely irregular and cannot be resolved into
independent subbands as in the case of the parameter set A. This is
a consequence of the mixing of the adiabatic reference systems and
the irregular structure of the eigenstates analyzed by the Husimi projections 
in the preceding section. 
In particular, these projections (see Fig.~\ref{fighus_B}) 
show a random variation of the intensity in the 
$Q=P=0$ region, which according to eq.(\ref{abshus}) is
relevant for the final states of the optical transitions.
This random variation is probed by optical transition matrix elements 
and results in an irregular pattern
of lines in the fine structure of the high energy absorption band. 

\begin{figure}
\centerline{\epsfig{file=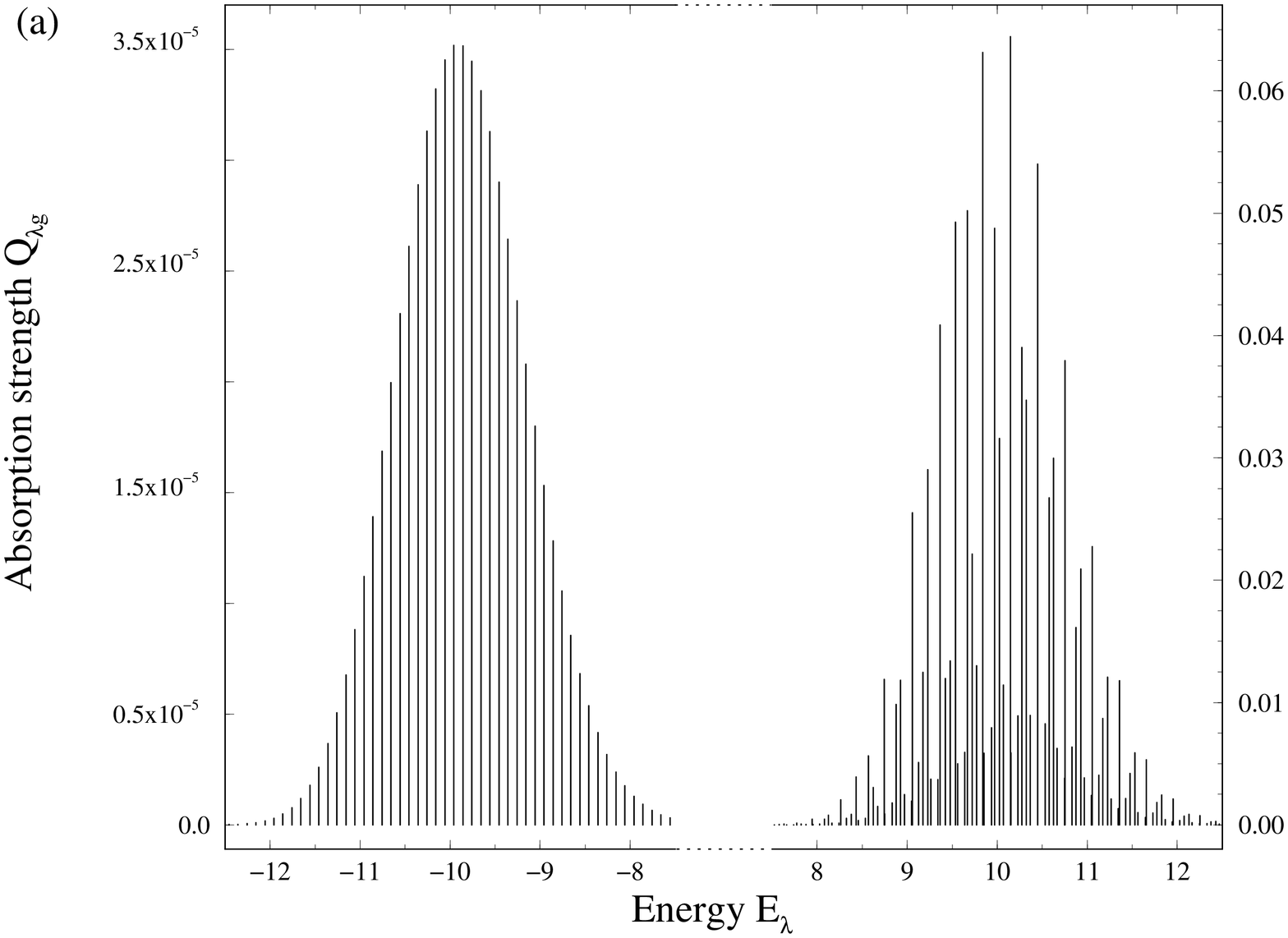,width=10cm,angle=270}}
\centerline{\epsfig{file=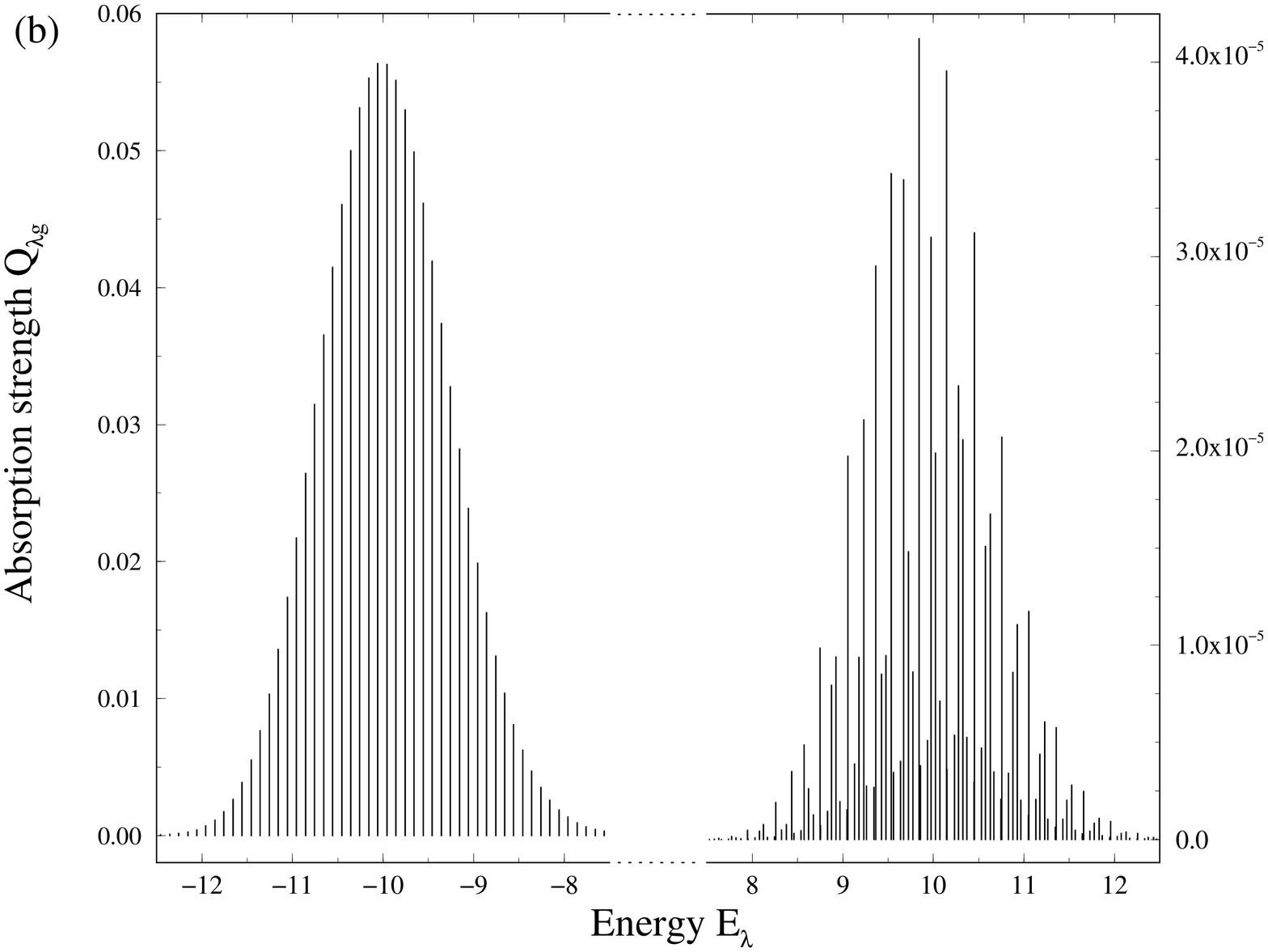,width=10cm,angle=270}}
\caption{
Absorption bands for the parameter set B
in the limiting cases of optical asymmetry, 
when one of the molecules of the dimer is
optically active only: 
$\mu_1 \neq0, \mu_2=0$ (shown in part (a)) and  
$\mu_2 \neq0, \mu_1=0$ (shown in part (b)).
Note the differences between the overall intensities 
of the lower and upper bands as indicated
by the scales on the left and right hand sides,
respectively: In (a) the intensity of the lower
band is by three orders of magnitude
smaller than the upper band, in (b) the
intensities of the bands are reversed.
Independent of this change in intensity
the upper bands in both (a) and (b) 
are irregular and have a similar 
fine structure. } 
\label{figabs_B2}
\end{figure}

\begin{figure}
\centerline{\epsfig{file=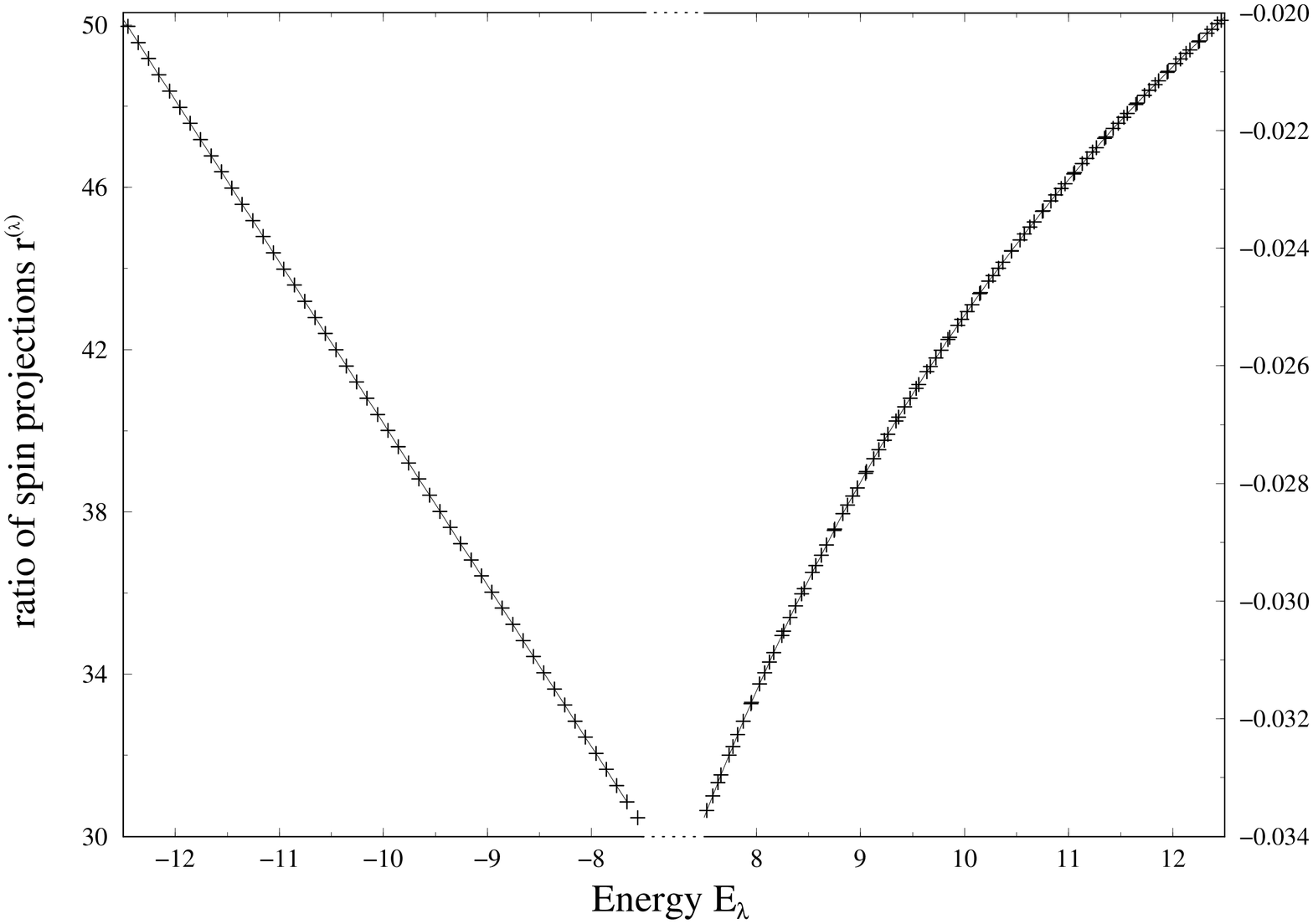,height=15cm,angle=270}}
\caption{
Ratio of the spin down to spin up coefficients 
$r^{(\lambda)}$ in the spectral regions 
of the lower and upper absorption bands 
of Fig.~\ref{figabs_B2}.
The scales on the left and right side 
correspond to the ratios in the 
region of the lower and upper band, respectively.
Note the smooth dependence of $r^{(\lambda)}$ on the
eigenstate energy in the spectral region of the upper
absorption bands, despite the irregular structure of the 
upper bands in Fig.~\ref{figabs_B1} and 
Fig.~\ref{figabs_B2} (a),(b).}
\label{figabs_r}
\end{figure}

The absorption bands in Fig.~\ref{figabs_A} and Fig.~\ref{figabs_B1} 
were calculated for equal optical transition matrix elements 
at the monomers $\mu_1=\mu_2$, which correspond to spin projections of the 
eigenstates with equal components. The changes occuring in the lower and
upper absorption bands in the limiting cases of an optical asymmetry 
with $\mu_1 \neq 0, \mu_2=0$ and $\mu_1=0, \mu_2 \neq 0$     
for the case of the parameter set B are compared in the 
Figs.~\ref{figabs_B2} (a) and (b), respectively. 
These limiting cases correspond to an optical 
asymmetry, when one of the monomers constituting the dimer is
optically active only. Then according to eq.(\ref{Melem})
the optical transition matrix element is determined by
either the spin up $c_{\uparrow 0}^{(\lambda)}$
or the spin down $c_{\downarrow 0}^{(\lambda)}$
coefficients of the $\lambda$ eigenstate.
As is seen from Fig.~\ref{figabs_B2},
which is representative for the case of a positive sign
of the excitation energy asymmetry $\epsilon_-$ 
($\epsilon_-=10>0$ in the case shown in Fig.~\ref{figabs_B2}),
the lower and upper bands 
show a redistribution of their intensities with 
the optical asymmetry: 
For the given sign of the excitation energy 
asymmetry $\epsilon_->0$ ($\epsilon_-<0$) 
the intensity of the
upper band is increased (decreased) compared 
to the symmetric case, $\mu_1=\mu_2$, displayed in Fig.~\ref{figabs_B1}. 
The lower band behaves in the opposite way,
the peak intensities of both bands differing by a factor 
of about $10^{-3}$. 
In particular, maximum absorption is reached for the high
energy band, when the monomer with the higher excitation energy is
optically active only,
i.e. $\epsilon_->0$ with $\mu_1\neq0$ and $\mu_2=0$.
Reversing the optical activity of the monomers but still considering
the same asymmetry in the excitation energies, 
i.e. $\epsilon_->0$ with $\mu_1=0$ and $\mu_2\neq0$,
one finds maximum absorption for the low energy band 
(if $\epsilon_-<0$, the quantities $\mu_1$ and $\mu_2$ have to be
interchanged in the above consideration). 
As in the case of equal transition matrix elements the lower
bands shown 
on the left sides of Fig.~\ref{figabs_B2}(a) and Fig.~\ref{figabs_B2}(b) 
display a regular fine structure, 
whereas the fine structure of the upper bands, displayed 
on the right sides of Fig.~\ref{figabs_B2}(a) and Fig.~\ref{figabs_B2}(b) 
is irregular. 
We note that the appearance of the weak bands is due to 
the coupling of the excited states of both monomers: 
The monomer, which is optically not active,
is still present in the absorption spectrum due to the coupling
of the excited states in the dimer system. 
In the case displayed 
on the right side of Fig.~\ref{figabs_B2}(b), e.g. 
an optically nonactive monomer with a high
excitation energy attached to an optically 
active monomer with a lower excitation energy produces
a weak but irregular band in the dimer system. 
A closer inspection of the fine structure 
of the irregular bands for the limiting cases of an optical asymmetry, 
in Fig.~\ref{figabs_B2}(a) and Fig.~\ref{figabs_B2}(b),
together with the irregular band for the symmetric case, shown
in Fig.~\ref{figabs_B1}, reveals that the sequence of lines with 
a strong and small absorption strength is almost identical in
all bands, there is only a small overall change 
in the line intensities in all three representations.
The reason behind this behavior becomes evident 
after an inspection of the ratio of the spin projections

\begin{equation}
\label{ratio}
r^{(\lambda)} = 
c_{\downarrow 0}^{(\lambda)} / c_{\uparrow 0}^{(\lambda)}, 
\end{equation}

which is presented in Fig.~\ref{figabs_r} as a function
of the eigenstate energy $E_{\lambda}$ for the spectral regions
corresponding to the lower and upper absorption bands, respectively.
As is evident from Fig.~\ref{figabs_r} 
$r^{(\lambda)}$ is a smooth function of the energy $E_{\lambda}$ 
in the spectral regions of both bands. In particular, for the case of 
the irregular upper absorption bands, for which in a sequence of 
eigenstates the spin projection coefficients 
$c_{\uparrow 0}^{(\lambda)}$ and $c_{\downarrow 0}^{(\lambda)}$ 
vary in an irregular way, their ratio $r^{(\lambda)}$ 
considered between neighbouring eigenstates is practically identical. 
This correlation between the spin up and spin down 
($c_{\uparrow 0}^{(\lambda)}$ and $c_{\downarrow 0}^{(\lambda)}$) 
coefficients results in an almost smooth interpolation
of the fine structure of the absorption bands between the limiting cases
displayed in the Fig.~\ref{figabs_B2}.
The smoothness of the function $r^{(\lambda)}$
can be used to express the absorption strength for arbitrary optical
asymmetry  by one of the limiting cases. 
For example, representing the
down coefficients $c_{\downarrow 0}^{(\lambda)}$ 
by the up coefficients $c_{\uparrow 0}^{(\lambda)}$ 
in (\ref{Melem}) using eq. (\ref{ratio}), one obtains 

\begin{equation}
\label{Qasy}
Q_{\lambda g} 
= \frac{(\mu_1 + \mu_2 r^{(\lambda)})^2}{\mu_1^2 + \mu_2^2}Q^{+}_{\lambda g},
\end{equation}

where $Q^{+}_{\lambda g}=[c_{\uparrow 0}^{(\lambda)}]^2$ is the
absorption strength for the case $\mu_1\neq0$ and $\mu_2=0$.
The smoothness of the ratio of 
the different spin projections $r^{(\lambda)}$ together with
eq. (\ref{Qasy}) allows an interpolation between absorption bands
for different optical asymmetries. In particular, using this ratio 
in the prefactor in front of $Q^{+}_{\lambda g}$ the relative 
intensities of absorption bands corresponding to different 
optical asymmetries can be determined. 

\section{Conclusions}

The transition from the coexistence to a mixing of the adiabatic branches
in the eigenstates of a symmetry-broken spin-boson Hamiltonian 
can be controlled by Bloch and Husimi projections and 
it shows up in the structure of absorption bands in 
excitonic-vibronic coupled dimer systems
with asymmetric adiabatic potential configurations.
In the mixing region of the adiabatic branches
phase space distributions of the eigenstates are 
concentrated on the phase space orbits of both adiabatic
potentials and their intensity varies in a random and irregular way.
This random variation in the intensitiy distribution 
of the eigenstates is probed by the optical 
transition matrix elements and it is a source of spectral randomness 
in the fine structure of the absorption bands.
The asymmetry of the optical transition matrix 
elements of the dimer configuration is
representable by suitably chosen spin projections of
the spin part of the eigenstates.
In the limiting cases of the optical asymmetry,
in which one of the monomers constituting the dimer
is optically active only, the intensities
of the absorption lines are determined by either the spin up or
spin down coefficients. For the states relevant for the
optical absorption correlated spin down and spin up 
coefficients are obtained. As a result the shape of 
the fine structure of the irregular absorption bands remains 
almost the same for different optical asymmetries.  
Finally we point out that in the case of a symmetry-broken
spin-boson Hamiltonian optical absorption processes
can probe the high energy regions of the overlap of 
adiabatic potentials where it is possible to observe
a transition from regular to irregular absorption
spectra for asymmetric molecular dimer configurations.

\newpage

\newpage

\newpage

\begin{thebibliography}{}
 
\bibitem{MSLN}
L. M\"uller, J. Stolze, H. Leschke, and P. Nagel, 
Phys. Rev. A 55, 1022 (1991).
\bibitem{SE97}
H. Schanz, and B. Esser, 
Phys. Rev. A 44, 3375 (1997).
\bibitem{ST}
G. Stock, and M. Thoss,
Phys. Rev. Lett. 78, 578 (1997). 
\bibitem{SSKR}
R. Steib, J. L. Schoendorff, H. J. Korsch, and P. Reineker, 
Phys. Rev. E 57, 6534 (1998).
\bibitem{BH}
J. S. Briggs, and A. Herzenberg,
Molecular Physics 23, 203 (1972).
\bibitem{SEW} 
M. Sonnek, H. Eiermann, and M. Wagner, 
Phys. Rev. B 51, 905 (1995).
\bibitem{RMB}
U. Rempel, B. von Maltzan, and C. von Borczyskowski,
J. Lumin. 53, 175 (1992).
\bibitem{T} 
K. Takahashi, 
Progr. Theor. Phys. Suppl. 98, 109 (1989).
\bibitem{LB}
W. A. Lin, and L. E. Ballentine,
Phys. Rev. Lett. 65, 2927 (1990)
\bibitem{CCLW}
M. B. Cibils, Y. Cuche, P. Leboef, and W. F. Wreszinski,
Phys. Rev. A 46, 4560 (1992).
\bibitem{K} 
M. Kus,
Phys. Rev. Lett. 54, 1343 (1985).
\bibitem{CCM} 
M. Cibils, Y. Cuche, and G. M\"uller, 
Z. Phys. B 97, 565 (1995). 
\bibitem{SE96} 
H. Schanz,  and B. Esser, 
Z. Phys. B 101, 299 (1996).
\bibitem{ZKCPD}
Th. Zimmermann, H. K\"oppel, L. S. Cederbaum, G. Persch, and W. Demtr\"oder,
Phys. Rev. Lett. 61, 3 (1988) 

\end{thebibliography}
\end{document}